\documentclass[prd,aps,twocolumn,preprintnumbers, showpacs, nofootinbib,superscriptaddress,notitlepage]{revtex4-1}
\usepackage{amssymb,amsthm,amsmath}
\usepackage{graphicx}   
\usepackage{color}      
\usepackage{slashed}    
\usepackage{verbatim}
\usepackage[normalem]{ulem}
\usepackage{rotating}   
\usepackage{multirow}   
\usepackage{subfigure}  
\usepackage{ulem}

\begin{document}

\title{Exploring Hidden-charm and hidden-strange Hexaquarks states from Lattice QCD}

\author{Hang Liu}
\affiliation{INPAC, Key Laboratory for Particle Astrophysics and Cosmology (MOE),  Shanghai Key Laboratory for Particle Physics and Cosmology, School of Physics and Astronomy, Shanghai Jiao Tong University, Shanghai 200240, China}

\author{Jinchen He}
\affiliation{INPAC, Key Laboratory for Particle Astrophysics and Cosmology (MOE),  Shanghai Key Laboratory for Particle Physics and Cosmology, School of Physics and Astronomy, Shanghai Jiao Tong University, Shanghai 200240, China}

\author{Liuming Liu}
\affiliation{Institute of Modern Physics, Chinese Academy of Sciences, Lanzhou, Gansu Province 730000, China}
\affiliation{University of Chinese Academy of Sciences, Beijing 100049, China}

\author{Peng Sun}
\affiliation{Institute of Modern Physics, Chinese Academy of Sciences, Lanzhou, Gansu Province 730000, China}
\affiliation{Department of Physics and Institute of Theoretical Physics,
Nanjing Normal University, Nanjing, Jiangsu 210023, China}

\author{Wei Wang}
\affiliation{INPAC, Key Laboratory for Particle Astrophysics and Cosmology (MOE),  Shanghai Key Laboratory for Particle Physics and Cosmology, School of Physics and Astronomy, Shanghai Jiao Tong University, Shanghai 200240, China}

\author{Yi-Bo Yang}
\affiliation{CAS Key Laboratory of Theoretical Physics, Institute of Theoretical Physics, Chinese Academy of Sciences, Beijing 100190, China}
\affiliation{School of Fundamental Physics and Mathematical Sciences, Hangzhou Institute for Advanced Study, UCAS, Hangzhou 310024, China}
\affiliation{International Centre for Theoretical Physics Asia-Pacific, Beijing/Hangzhou, China}
\affiliation{School of Physical Sciences, University of Chinese Academy of Sciences,
Beijing 100049, China}

\author{Qi-An Zhang}
\affiliation{
School of Physics, Beihang University, Beijing 102206, China}

\begin{abstract}
Based on five different ensembles of newly-generated  (2+1)-flavor configurations with the pion mass around $m_{\pi}{\simeq} (140-310)$ MeV,  we present a lattice analysis of hidden-charm and hidden strange hexaquarks with the quark content $usc\bar{d}\bar{s}\bar{c}$.
The correlation matrix of two types of operators with  $J^{PC}=0^{++}, 0^{-+}, 1^{++}$ and $1^{--}$  are simulated to extract the masses of hexaquark candidates which are then extrapolated to the physical pion mass and the continuum limit. Results indicate that masses of the ground states are below the $\Xi_c \bar \Xi_c$ threshold and provide a characteristic signal for the experimental discovery of hexaquark candidates. This may enrich the versatile structures of  multiquarks and is an indispensable step to decipher the nonperturbative nature of fundamental interactions of quarks and gluons. 
\end{abstract}
\maketitle

{\it Introduction:} 
The spectrum of hadron  excitations discovered  at experimental  facilities around the world manifests the fundamental interactions of elementary quarks and gluons, governed by the quantum gauge field theory of QCD. Understanding the complex emergent phenomena of this field theory has captivated the attention of theoretical physicists in the last decades. To date one of the perennial problems in hadron physics is to establish  the existence of exotic hadrons that defy the quark-antiquark interpretation for mesons  and three-quark assignment for baryons~\cite{Gell-Mann:2015moa,Zweig:1964jf,Jaffe:1976ig,Jaffe:1977cv,Lipkin:1987sk}. Candidates of such exotic hadrons,  including  tetraquark and pentaquark states,  have been recently discovered and confirmed in various experimental measurements~\cite{Belle:2003nnu,CDF:2003cab,D0:2004zmu,LHCb:2014zfx,BESIII:2016adj,BESIII:2016bnd,BESIII:2017tqk,BaBar:2005hhc,CLEO:2006tct,Belle:2007dxy,Belle:2013yex,BaBar:2006ait,BaBar:2012hpr,Belle:2007umv,Belle:2014wyt,BESIII:2020qkh,LHCb:2015yax,LHCb:2016ztz,LHCb:2019kea,LHCb:2020bwg,CMS:2022yhl,ATLAS:2022hhx}. These experimental progress give us strong  confidence for the existence of  hexaquark states.

The proposition of six quarks  as  a single hadron structure  was first presented  in 1964 \cite{Dyson:1964xwa}, and a renowned  realization is deuteron. In addition,  the $d^*(2380)$ resonance, reported by CELSIUS/WASA and WASA-at-COSY collaborations \cite{Faldt:2011zv,WASA-at-COSY:2011bjg,WASA-at-COSY:2012seb,Kim:2020rwn,Dong:2018emq},  is widely believed to be a dibaryon.  Until now, a lot of explorations of  hexaquarks with different quark constituents have been conducted in theory, such as heavy dibaryons (qqqqqQ)~\cite{Oka:2019mrd,Pepin:1998ih}, doubly-heavy dibaryons (qqqqQQ)~\cite{Vijande:2016nzk,Meng:2017fwb}, fully light dibaryons~\cite{Zhang:1997ny,  Gerasyuta:2010hn, Park:2015nha, Chen:2019vdh}, and full heavy dibaryons~\cite{Huang:2020bmb}. In view of the {\it ab-initio} framework--lattice QCD, the major challenges include not only the accurate extraction of the bounding energy~(E.g.,\cite{NPLQCD:2012mex}), but also the complicated contraction involving many quarks with the same flavor.

On the other hand, hadrons composed of three quarks and three anti-quarks belong to another category of hexaquarks. Other than light ones, hidden-charm and hidden-bottom hexaquarks are of special interest since heavy quarks have much larger masses and thus are more easily distinguished from ordinary mesons. Investigating this type of hexaquarks through lattice QCD is even harder than the dibaryons, due to the mixing with the three meson states.

In this work, we show a lattice QCD investigation of the hadronic structures containing three quarks and three antiquarks, both using the bayron-anti-bayron type and three-meson type interpolation fields with $I(J^{PC})=1(0^{++}),1(0^{-+}),1(1^{++}),1(1^{--})$. To minimize the impact of disconnected diagrams, we have opted for a quark composition of $usc\bar{d}\bar{s}\bar{c}$ in the case of isospin $I=1$. In this case, the annihilation diagrams of charm and strange quarks may contribute, but their contribution is expected to be suppressed by the OZI rule and therefore are ignored in the first-step study. After making the chiral and continuum extrapolation, we find that the spectrum of hexaquarks which are below the $\Xi_{c} \bar \Xi_{c}$ threshold, and two ground states are close to the three-meson thresholds. This feature is consistent with the result from the chromo-magnetic interaction model which also finds a large binding energy for the hexaquark states with certain quantum numbers~\cite{Liu:2021gva}, but quite different from the model predictions which focus on the near-threshold structure of $\Xi_{c} \bar \Xi_{c}$ only. This interesting observation can be further examined by more theoretical analyses and validated by future experimental measurements.

{\it Theoretical Framework:} 
A most powerful method to systematically  tackle the nonperturbative strong interaction is Lattice QCD~\cite{Wilson:1974sk}, in which the quark and gluon ﬁelds are discretized on a space-time grid of ﬁnite size, allowing numerical computation by averaging over large numbers of possible ﬁeld conﬁgurations generated by Monte-Carlo. In particular, from the time dependence of correlation functions calculated in this way, one can determine a discrete spectrum of various hadrons.  Thus  Lattice QCD provides a first-principles technique for explorations of quantities of interest, such as spectrum, scattering phase, and radiative transitions~\cite{Dudek:2007wv,Dudek:2006ej,Dudek:2009kk}.

 To determine the mass spectrum, one firstly needs to construct appropriate interpolating operators with definite symmetries. We use the baryon-anti-baryon type (denoted by superscript A) and three-meson type (denoted by superscript B) interpolation fields to construct our correlation function matrix. Therefore, for the hexaquarks, one can construct the interpolating operators with quantum numbers $0^{++},0^{-+},1^{++},1^{--}$ as

\begin{eqnarray}
0^{++}:\notag\\
\mathcal{O}^A_1(x)&=&\epsilon^{abc}\epsilon^{def}[{u^{T}_{a}}C\gamma_5{s_{b}}][{\bar{d}_{d}}C\gamma_5{\bar{s}_{e}}^{T}]\times  [{\bar{c}_{f}}c_{c}](x), \notag \\
\mathcal{O}^B_1(x)&=&[\bar{s}\gamma_5u]\times [\bar{d}\gamma_5 s] \times[ \bar{c}c](x),\notag \\
0^{-+}:\notag\\
\mathcal{O}^A_2(x)&=&\epsilon^{abc}\epsilon^{def}[{u^{T}_{a}}C\gamma_5{s_{b}}][{\bar{c}_{d}}C\gamma_5{\bar{s}_{e}}^{T}] \times  [{\bar{c}_{f}}\gamma_5c_{c}](x), \notag \\
\mathcal{O}^B_2(x)&=&[\bar{s}\gamma_5u]\times [\bar{d}\gamma_5 s] \times[ \bar{c}\gamma_5 c](x),\notag \\
1^{++}:\notag\\
\mathcal{O}^A_3(x)&=&\epsilon^{abc}\epsilon^{def}[{u^{T}_{a}}C\gamma_5{s_{b}}][{\bar{d}_{d}}C\gamma_5{\bar{s}_{e}}^{T}]\times  [{\bar{c}_{f}}\gamma_i\gamma_5 c_{c}](x), \notag \\
\mathcal{O}^B_3(x)&=&[\bar{s}\gamma_5u]\times [\bar{d}\gamma_5 s] \times[ \bar{c}\gamma_i \gamma_5 c](x),\notag \\
1^{--}:\notag\\
\mathcal{O}^A_4(x)&=&\epsilon^{abc}\epsilon^{def}[{u^{T}_{a}}C\gamma_5{s_{b}}][{\bar{d}_{d}}C\gamma_5{\bar{s}_{e}}^{T}] \times  [{\bar{c}_{f}}\gamma_i c_{c}](x),\notag\\
\mathcal{O}^B_4(x)&=&[\bar{s}\gamma_5u]\times [\bar{d}\gamma_5 s] \times[ \bar{c}\gamma_i c](x).
\label{eq:operators}
\end{eqnarray}
Here $C=i\gamma_2\gamma_4$ is the charge conjugation matrix and $a,...,f$ are color indices. The operators at the source on the Coulomb gauge fixed configuration would be the following two kinds:
\begin{align}
O^{A}_{(s)}(t)&=\sum_{\vec{y}_i,i=1,6} \epsilon^{abc}\epsilon^{def}[{u^{T}_{a}(\vec y_1)}C\gamma_5{s_{b}(\vec y_2)}]\notag\\
&[{\bar{d}_{d}(\vec y_3)}C\gamma_5{\bar{s}_{e}(\vec y_4)}^{T}] \times  [{\bar{c}_{f}(\vec y_5)}\gamma_x c_{c}(\vec y_6)],\\
O^{B}_{(s)}(t)&=\sum_{\vec{y}_i,i=1,6} \bar{s}(\vec y_1)\gamma_5 u(\vec y_2)\times\bar{d}(\vec y_3) \gamma_5 s(\vec y_4)  \notag\\
&\times\bar{c}(\vec y_5) \gamma_x c(\vec y_6),
\end{align}
where $\gamma_x$ can be $1,\gamma_5,\gamma_i\gamma_5,\gamma_i$ corresponding to quantum numbers $0^{++},0^{-+},1^{++},1^{--}$ and all six positions are integrated separately, as we are using the Coulomb wall source. The operators at the sink are using only one position integration. It should be noticed that there are various potential operators that can be used, and a comprehensive treatment should take into account all these operators, and simulate the corresponding correlation functions.  In this work,  we have opted for baryon-anti-baryon interpolating operators in the form of $\Xi^+_c\bar{\Xi}^0_c$, and include the corresponding three-meson interpolating operators represented by $K\bar{K}\eta_c$ of quantum numbers $0^{-+}$ ($K\bar{K} J/\psi$ of quantum numbers $1^{--}$ and so on) which might give large contributions to the correlation functions.

Then the determination of the mass spectrum proceeds from the calculation of correlation functions matrices between this operator and  its hermitian conjugate at Euclidean times $t$ and $0$ of the form
\begin{align}
C^{\alpha\beta}_i(t)=\langle 0|\mathcal{O}^\alpha_i(t)\mathcal{O}^{\beta\dagger}_{i(s)}(0) |0\rangle, 
\label{eq:2pt}
\end{align}
where  $i$ labels operators with the four different quantum numbers, and $\alpha,\beta$ can be either $A$ or $B$. For each quantum number, we evaluate
a $2\times 2$ correlation matrix and then we solve the equation for the generalized eigenvalue problem(GEVP)~\cite{Dudek:2010wm,Blossier:2009kd}:
\begin{align}
C(t)v_n(t,t_0)=\lambda(t,t_0)C(t_0)v_n(t,t_0)
\end{align}
where $t_0$ is a reference time slice, $\lambda$ is the eigenvalue of the matrix $C(t_0)^{-1}C(t)$ and $v_n$ being the eigenvectors correspondingly. Normally one chooses $t_0$ large enough and the signal is good and stable. The parameter $t_0$ is tunable
and one could optimize the calculation by choosing $t_0$ such that the correlation matrix is dominated by the desired eigenvalues at that particular $t_0$ (preferring a larger $t_0$) with an acceptable signal-to-noise ratio (preferring a smaller $t_0$). The energy
eigenvalues for the system are then obtained
by diagonalizing the matrix $C(t_0)^{-1}C(t)$(or $C(t)C(t_0)^{-1}$). The eigenvalues of
the matrix have the usual exponential decay behavior as
described by Eq.~\eqref{eq:lambdan} and therefore the exact energy $E_n$
can be extracted from the effective mass plateau of the
eigenvalue $\lambda_n$.
\begin{align}
\lambda_n(t,t_0)=e^{-E_n(t-t_0)}(1+\mathcal{O}(e^{-|\delta E|(t-t_0)}))
\label{eq:lambdan}
\end{align}
where $\delta E$ is the energy gap between $E_{n+1}$ and $E_n$. Including only correlation functions projected to zero momentum, we have $E_n = M_n$, which yields the ground state. The effective masses can be obtained from two-state fits of the eigenvalues or the plateau fit of the effective masses. 
 \hspace*{\fill} \\

{\it Lattice Simulation:}~We employ (2+1)-flavor Wilson clover fermion gauge configurations generated with the lattice spacings  $a=0.054\rm fm$, $0.080 \rm fm$, $0.108\rm fm$. A first analysis of $\Xi_c\to \Xi$ form factors using two ensembles  (C08P30S and C11P29S) has been conducted in Ref.~\cite{Zhang:2021oja}, and predictions on partial widths for semileptonic $\Xi_c$ decays were used in the experimental background simulation by Belle collaboration~\cite{Belle:2021crz}.   Tab.~I shows the parameters of these configurations. The pion masses and the lattice spacings are given in units of $\rm MeV$ and $\rm fm$, respectively. The bare strange quark mass is determined such that the mass of $\eta_s$ is around $700$ MeV~\cite{MILC:2010pul, Koponen:2017ayj},  and the bare charm quark mass is tuned to accommodate the spin-average value of the $J/\psi$ and $\eta_c$ masses ($\frac{1}{4}m_{\eta_c}+\frac{3}{4}m_{J/\psi}=3.069\rm{GeV}$). The quark propagators are computed using the Coulomb gauge fixed wall source.


\begin{table}[http]
\begin{center}
\begin{tabular}{c|cccccccccc}
  & $\beta$ & $L^3\times T$ & $a$  &$\kappa_l$ & $m_{\pi}$ & $\kappa_s$  & $\kappa_c$&\\\hline
 C11P29S&$6.2$ & $24^3 \times 72$ &$0.108$ &$0.1343$ &$284(2)$ & $0.1327$   & $0.1117$\\\hline
 C11P22M&$6.2$ & $32^3 \times 64$ &$0.108$  &$0.1344$ &$221.8(7)$ & $0.1326$   & $0.1116$  \\\hline
 C11P14L&$6.2$ & $48^3 \times 96$ &$0.108$  &$0.1345$ &$135.1(6)$ & $0.1327$   & $0.1116$  \\\hline 
 C08P30S&$6.41$ & $32^3 \times 96$ &$0.080$ &$0.1326$ &$296.8(9)$ & $0.1316$   & $0.1181$ \\\hline
 C06P30S& $6.72$ & $48^3 \times 144$ & $0.054$ & $0.1311$ & $312(1)$ & $0.1305$ & $0.1227$ \\
\end{tabular}
\caption{\label{tab:widgets}Parameters of the 2+1 flavor Wilson clover fermion ensembles used in this calculation. The pion  masses and the lattice spacings are given in units of MeV, and fm, respectively.   }
\end{center}
\end{table} 
We first analyze the ordinary mesons and baryons including the $\pi$, $K$,  $D$, $D_s$,  $\Lambda$,  $\Xi$, $\Omega$, $\Lambda_c$, $\Xi_c$, $\eta_c$ and $J/\psi$.  Using the wall sources with Coulomb gauge fixing, we calculate the two-point correlation functions with the pertinent interpolating operator. 
The mass for the hadrons can be obtained  from two-state fits of the correlations. To explore the effects of the finite lattice spacing and non-physical pion mass, we perform a joint extrapolation to the continuum and physical pion limit. We extrapolate all of the ground state masses to the physical pion mass and continuum limit using the following parametrization for all   the hadrons except kaon~\cite{Gasser:1984gg},
\begin{eqnarray}
 m_H(m_\pi, a) = m_{H, {\rm phys}} + g^H_1(m_\pi^2 {-m_{\pi, {\rm phys}}^2}) + g^H_2 a^2,
 \label{eq:joint-extrapolate}
\end{eqnarray}
and the parameterization inspired by the chiral symmetry is used in the kaon case:
\begin{eqnarray}
 m_K^2(m_\pi, a) =m_{K, {\rm phys}}^2 + g^K_1 (m_\pi^2 {-m_{\pi, {\rm phys}}^2}) + g^K_2 a^2,
\end{eqnarray}
where $m_{\pi, {\rm phys}}=135$~MeV is the physical pion mass without iso-spin symmetry breaking and QED effect. In the extrapolation, the statistical errors in each ensemble are taken as independent, and the final errors are obtained after making the extrapolation. Systematic uncertainties and those uncertainties from the lattice spacing of the configurations are not included.  As shown in TABLE.II, all our predictions after the continuum extrapolation and chiral extrapolation (to physical pion-mass) agree with experimental values within  {2$\sigma$ deviation}, except the $\Xi_c$ and $\eta_c$ masses which require better control on the pion-mass extrapolation, charm quark mass determination and systematic error allowance\\
\begin{table}[hbp]
\begin{center}
\begin{tabular}{c|cccccccccc}
\hline
   Hadron    & $K$ &$D$ & $D_s$ & $\Lambda$   \\\hline
Lattice    & 0.4869(41)  & 1.8675(76) & 1.9766(65)   & 1.074(48) 
  \\\hline
 Exp.   &0.4937/0.4976   &1.864/1.870 &$1.968$ & 1.115    \\\hline 
Hadron   &   $\Xi$ & $\Omega$ & $\Lambda_c$ &  $\Xi_c$ & \\\hline
Lattice    &1.354(22)   &1.699(42) & 2.348(59)   &2.4380(68) & 
  \\\hline
 Exp.   &$1.314$  &$1.672$ &$2.286$ & $2.468$     \\\hline 
Hadron   &   $\eta_c$ & $J/\psi$ &  \\\hline
Lattice    & 3.0041(20)& 3.0972(24)
  \\\hline
 Exp.   & 2.9839  & 3.0969    \\\hline
\end{tabular}
\caption{ Mass (in the unit of GeV) for the ordinary hadrons.  With the five ensembles of configurations, we have extracted the mass from the analysis of two-point correlation functions and extrapolated to the continuum and physical pion mass limit, and the errors are statistical. The experimental data are taken from Particle Data Group~\cite{Zyla:2020zbs}.  }
\label{tab:ordinary_meson}
\end{center}
\end{table} 
Results for these states are collected in Tab.~\ref{tab:ordinary_meson}, in which we also collect the experimental measurements from Particle Data Group~\cite{Zyla:2020zbs}. From this table, one can notice that the newly-generated configurations can give a reasonable description of all these hadrons. More details about the calculation will be given in the following.

\begin{figure}[http]
\centering
\includegraphics[width=0.45\textwidth]{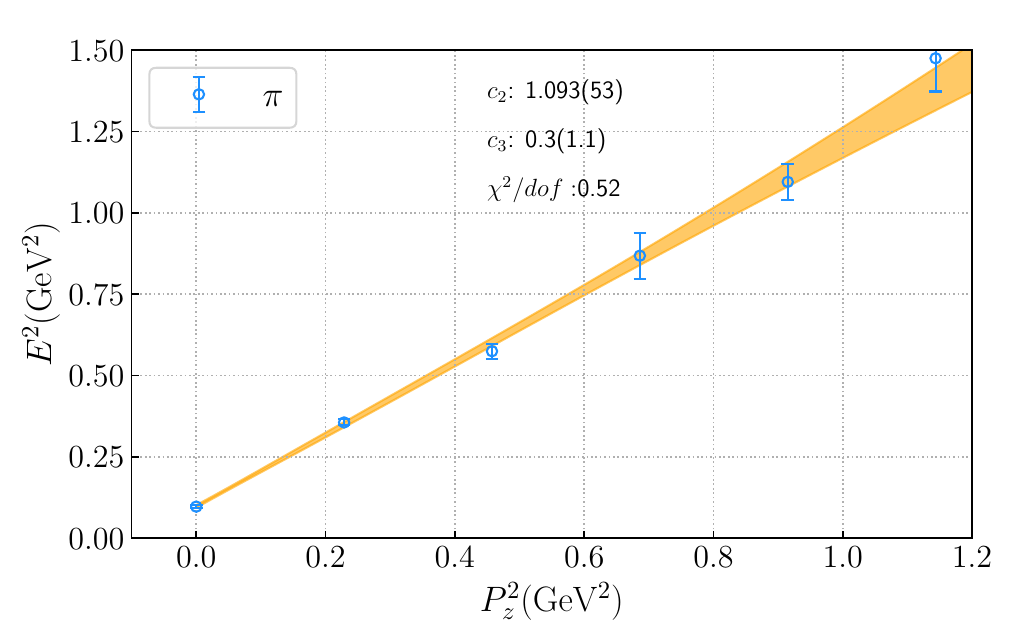} 
\includegraphics[width=0.45\textwidth]{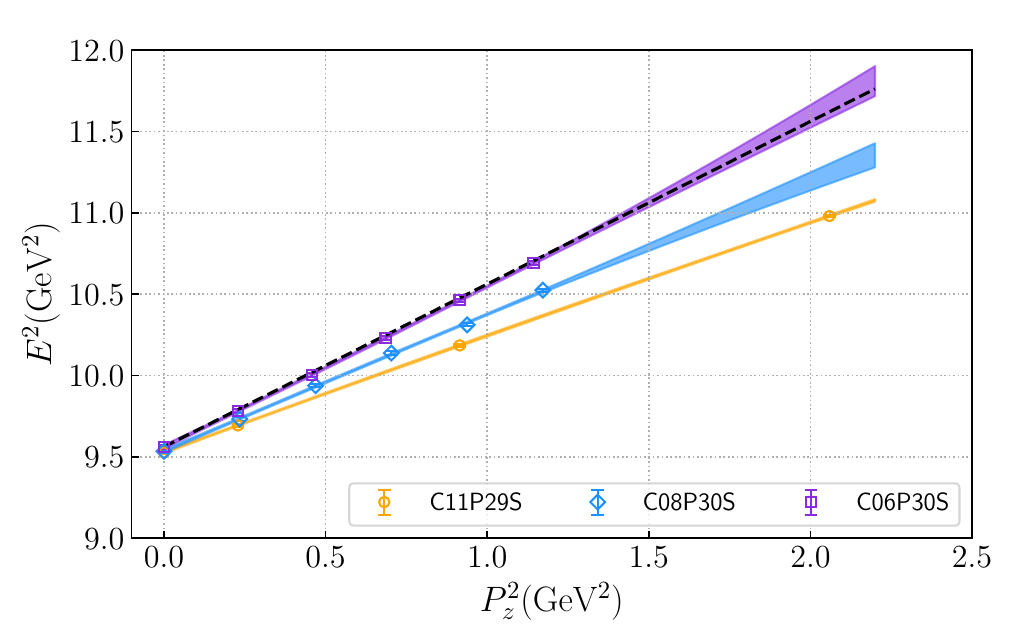}
\caption{\label{fig:dispersion-C06P30S}The dispersion relation for $\pi$ on the ensemble C06P30S (upper panel) and for $J/\psi$ on the three ensembles C11P29S, C08P30S, and C06P30S (lower panel). }
\end{figure}

In addition,  we  explore the  mass-energy dispersion relations for the ordinary mesons  to ensure that all the lattice discretization errors are under control. We have generated the two-point correlation functions  for the $\pi$, $\eta_s$, and $J/\psi$ mesons at different momenta, and obtained the energy from two-state fits of the correlations.  Some of the results are shown in Fig.~\ref{fig:dispersion-C06P30S}. The upper panel of Fig.~\ref{fig:dispersion-C06P30S} shows the dispersion relation for $\pi$ on the ensemble C06P30S where six momenta are chosen. Then the dispersion relation is parametrized as 
\begin{align}
  E^2=m^2+c_2 p^2+c_3 p^4 a^2,  \label{eq:disper}
\end{align}
where  $c_2$ and $c_3$ are parameters to be obtained through a fit. Deviations of $c_2$ from unity and $c_3$ from zero characterize the discretization errors. As one can see from Fig.~\ref{fig:dispersion-C06P30S}, all lattice results can be well described by Eq.~\eqref{eq:disper} with a reasonable $\chi^2/d.o.f$. 
The results $c_{2,\pi}=1.123(69)$, $c_{2,\eta_s}=1.064(47)$, $c_{2,J/\psi}=0.954(45)$  are consistent with the square of the speed of light, while the $c_3$ parameters are all close to 0. 
From the results, one can notice that the dispersion relations for the ordinary mesons with $u/d, s,c$ quarks are well-respected on these lattice configurations after the continuum extrapolation.

\begin{figure}[hbp]
\centering
\includegraphics[width=0.48\textwidth]{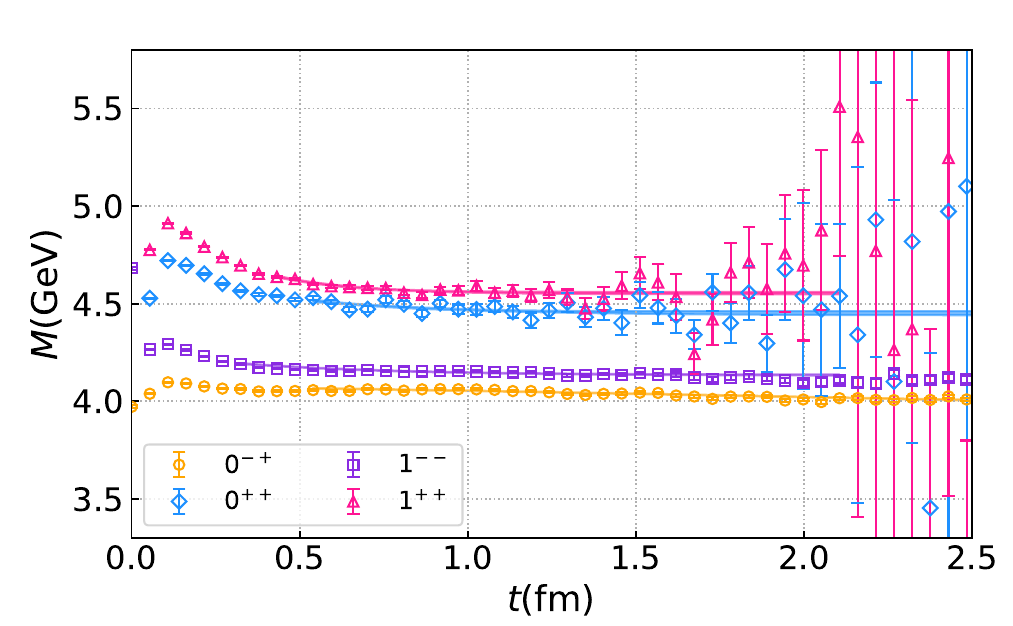}
\caption{Effective mass of the ground states for hexaquarks on the four different quantum numbers on C06P30S ensembles. The yellow markers and the corresponding fit line represent the effective mass of the ground state, derived from a $2\times2$ correlation matrix for the quantum number $0^{-+}$. The effective mass corresponds to the mass of the lower state, which is obtained through diagonalizing the $2\times 2$ matrix element. Conversely, the purple markers denote the effective mass of the ground state from the $2\times2$ correlation matrix for the quantum number $1^{--}$. In a similar manner, the blue ones for the quantum number $0^{++}$ and the pink ones for the quantum number $1^{++}$.}
\label{fig:effective_mass_C06P30S}
\end{figure}

{\it Numerical Results for Hexaquarks:}~The  focus of this work is  the states with $I=1,J^{PC}=0^{-+},0^{++},1^{--},1^{++}$. In the calculation of the two-point correlation functions matrix in Eq.~\eqref{eq:2pt},  we have  used $399\times20$, $451\times48, 203\times48,653\times 40$ and $136\times80$ "configurations$\times$loop-t" for C11P29S, C11P22M, C11P14L, C08P30S and C06P30S ensembles, respectively.  To extract the mass for hexaquark states, we adopt the  two-state parametrization for the eigenvalues obtained through diagonalizing the $2\times 2$ matrix element:
\begin{eqnarray}
\lambda(t,t_0)= e^{-M_H(t-t_0)} [1+ \Delta c\times  e^{-\Delta E (t-t_0)}]. 
\end{eqnarray}
where $\Delta c, M_H$ and $\Delta E$ are parameters to be determined through a correlated fit of the lattice data.   $M_H$ is the lowest-lying hexaquark state and $\Delta E$ corresponds to the relative mass gap of the excited state. It is necessary to stress that the $\Delta E$ is an effective energy reflecting the contributions from all possible higher states in the spectrum.   $\Delta c$ reflects the excited state contamination.  We perform two independent fits by fitting the correlation functions and the effective masses $M_{\rm eff}(t) = \ln (\lambda(t-1)/\lambda(t))$ and find consistent results for the $M_H$ within $1\sigma$.    As an illustration, the effective masses and the fitted energy plateaus for the
ensemble C06P30S are plotted in Fig.~\ref{fig:effective_mass_C06P30S}.
From the figures,  we find that all the lattice data can be well described with $\chi^2/d.o.f.$ smaller than 1.

\begin{figure}
\centering
\includegraphics[width=0.48\textwidth]{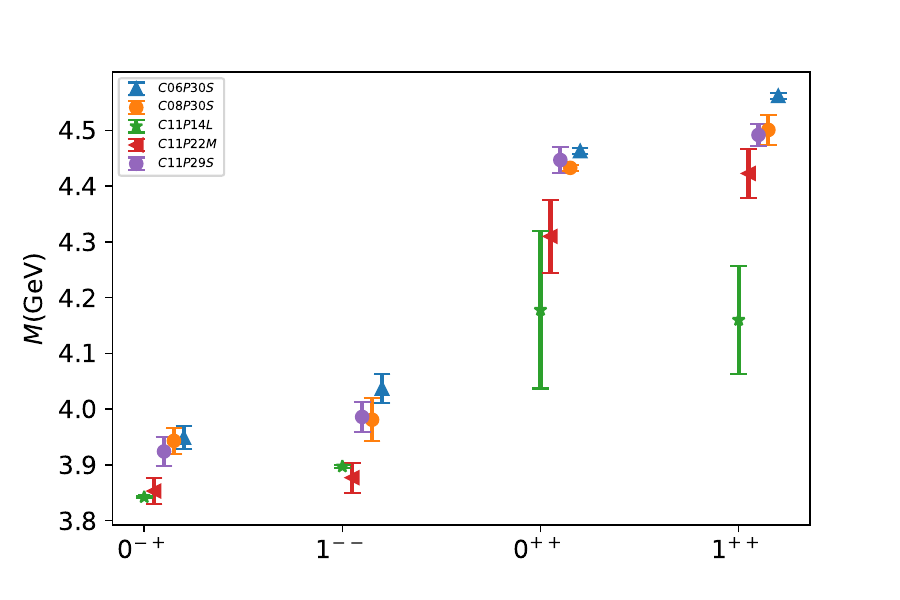}
\caption{The obtained masses of different quantum numbers on the five ensembles. The horizontal axis $0^{-+}$,$1^{--}$,$0^{++}$, $1^{++}$ are correspond to the corresponding ground states.}
\label{fig:spec_all}
\end{figure}

\begin{widetext}

\begin{figure}
\centering
\subfigure[~$0^{-+}$]{\label{fig:subfig:a}
\includegraphics[width=0.45\linewidth]{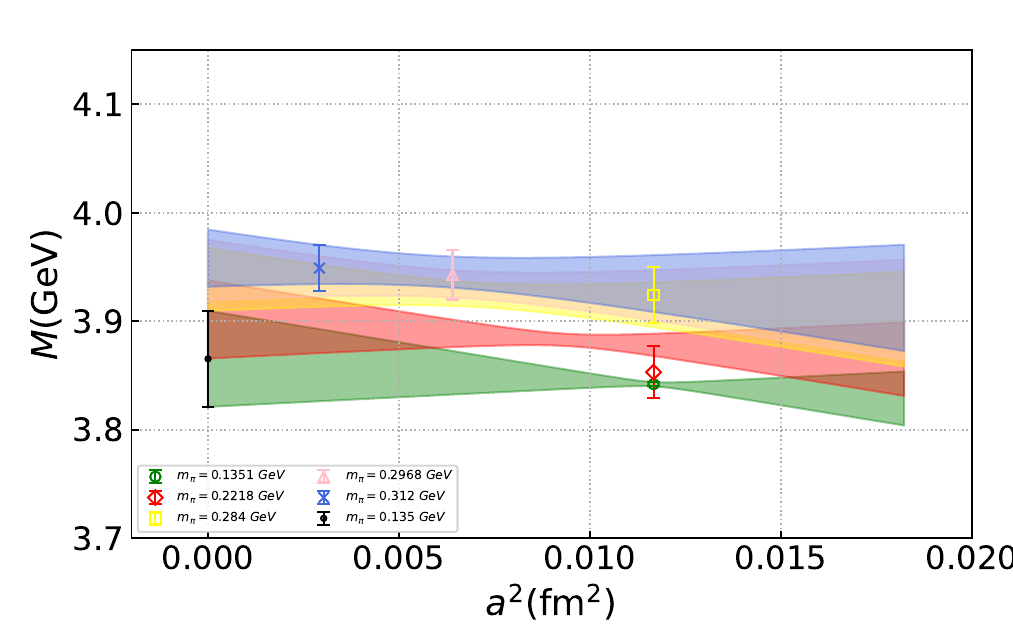}}
\hspace{0.01\linewidth}
\subfigure[~$0^{++}$]{\label{fig:subfig:b}
\includegraphics[width=0.45\linewidth]{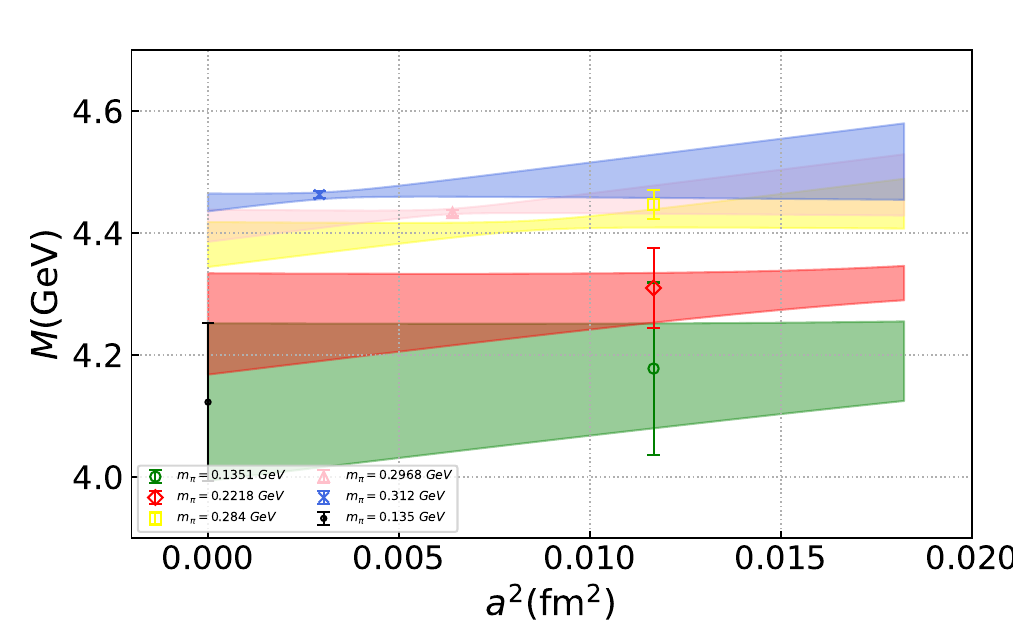}}
\vfill
\subfigure[~$1^{--}$]{\label{fig:subfig:a}
\includegraphics[width=0.45\linewidth]{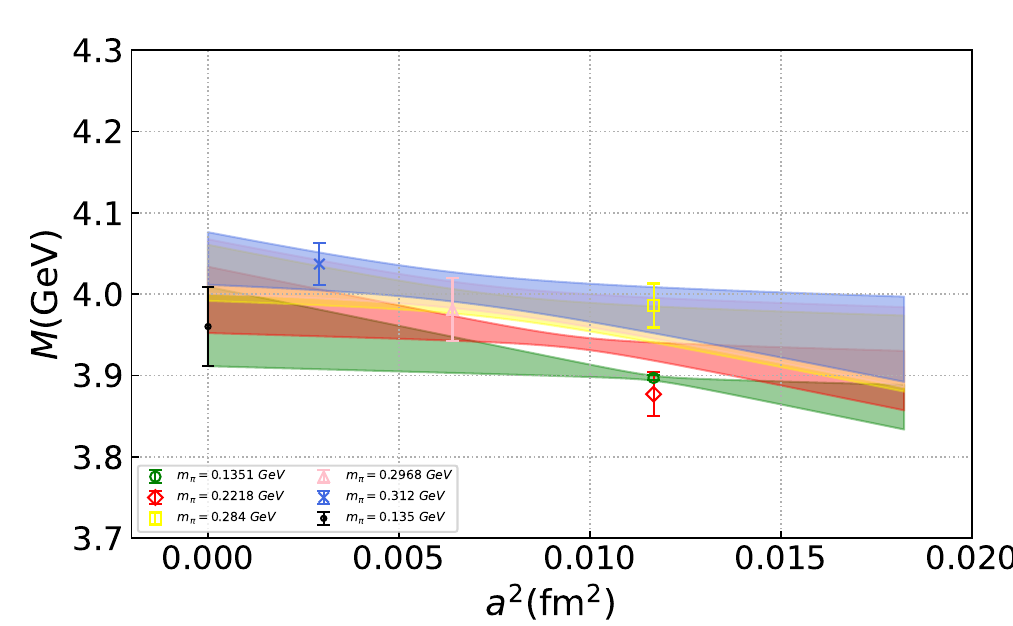}}
\hspace{0.01\linewidth}
\subfigure[~$1^{++}$]{\label{fig:subfig:b}
\includegraphics[width=0.45\linewidth]{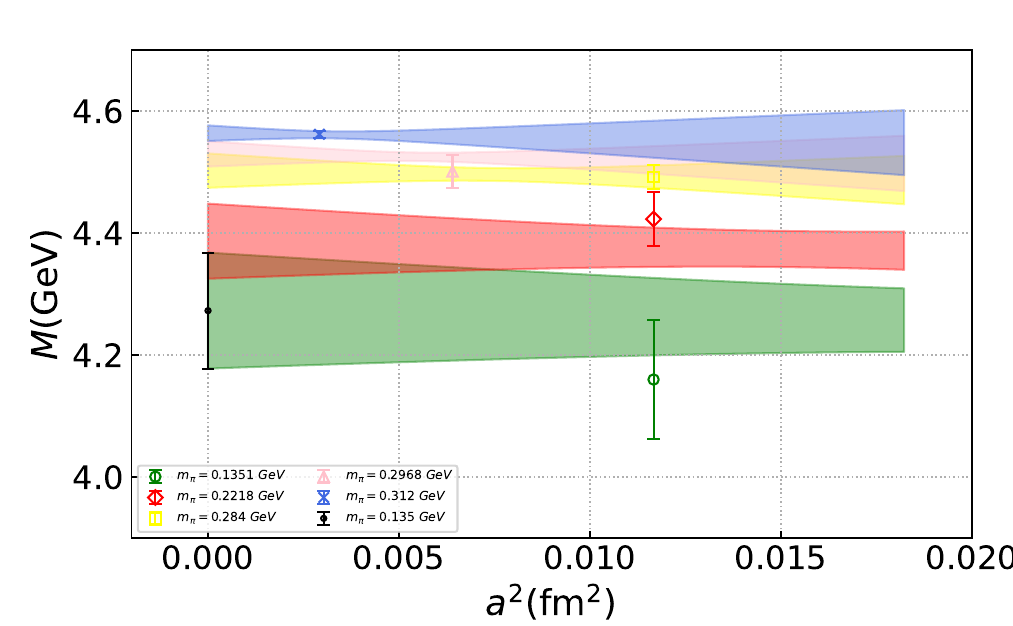}}
\caption{The continuum and physical pion mass extrapolation for the mass of hexaquark states.}
\label{fig:extrapolation}
\end{figure}

\end{widetext}

Results with the five ensembles for the masses are collected in Fig.~\ref{fig:spec_all}. To accommodate the effects caused by the nonphysical pion mass and discretization,  we also perform a simultaneous extrapolation of the masses for $m_{\pi}$ and lattice spacing $a$ using the parametrization as Eq.\eqref{eq:joint-extrapolate}.
The fit plots are 
 shown in Fig.~\ref{fig:extrapolation}, with a reasonable $\chi^2$ and the final results of the hexaquark states at the physical pion mass are shown in Tab.~\ref{tab:mass_chiral_extrapolation}.

\begin{table}[http]
\begin{tabular}{c|c|c|c|c}
\hline
$I(J^{PC})$ & $1(0^{-+})$ & $1(1^{--})$  &$1(0^{++})$ & $1(1^{++})$  \\\hline
  mass(GeV) &  {$3.865(44)$} &  {$3.960(49)$} &  {$4.12(13)$} &  {$4.273(95)$}\\
\hline
\end{tabular}
\caption{ The obtained hexaquark masses (in units of GeV) for the ground states at the physical pion mass after the chiral and continuum  extrapolation}
\label{tab:mass_chiral_extrapolation}
\end{table}

{\it Discussions:}~Our final results of the hexaquarks states with quantum numbers $0^{-+},1^{--},0^{++},1^{++}$ at the physical $m_{\pi}$ and at continuum limit are shown in the following Tab.~\ref{tab:mass_chiral_extrapolation}. 
It is interesting to notice that the threshold of $\Xi^{+}_c-\bar{\Xi}^0_c$ is about $4.938{\rm GeV}$~\cite{Zyla:2020zbs}, while the results for the hexaquark states are below this threshold by around 700 to 1000 MeV. This feature is consistent with the result from the chromo-magnetic interaction model which also finds a large binding energy for the hexaquark states with certain quantum numbers~\cite{Liu:2021gva}. 
But it should be noticed that a more conclusive statement requests more studies by including more channel effects. In addition the experimental mass summation of three mesons $K^+$, $K^-$ and $\eta_c$ is $3.971$ GeV~\cite{Zyla:2020zbs}, and according to Tab.~\ref{tab:mass_chiral_extrapolation} the mass for the $1(0^{-+})$ state  is slightly below this threshold. We note that the threshold of three mesons $\pi^+$, $\eta$ and $\eta_c$ is $3.667$ GeV and the one of three mesons $\pi^+$, $\eta'$ and $\eta_c$ is $4.077$ GeV. The mass for the $1(0^{-+})$ state is above the threshold of three mesons $\pi^+$, $\eta$ and $\eta_c$, it is also likely that the lowest state of hexaquark shows signals of a three-meson state $\pi \eta \eta_c$ with relative nonzero angular momenta. Since in this analysis the disconnected diagrams are not considered,  the constituent $\bar{u}u/\bar{d}d$ of $\eta(\eta')$ states within the spectrum is likely to be suppressed. As a result,   the particle combination of the
lowest state with $J^{PC}=0^{-+}$ might be  $\pi\eta_s\eta_c$, since the unphysical $\eta_s$ dominates the contribution to the
connected strange quark diagram in the pseudoscalar channel. An experimental search in this energy would be very helpful to clarify this finding. Theoretically more extensive studies on the hexaquark spectra by including more interpolating operators are required to further clarify the properties of the hexaquarks. We defer these studies to the future.

{\it Conclusions:}
Based on the newly generated  (2+1)-flavor configurations with the pion mass around $m_{\pi}{\simeq}140-310$ MeV,  we have presented a first lattice simulation of hidden-charm and hidden-strange hexaquark states with the quark content $usc\bar{d}\bar{s}\bar{c}$.  Four different quantum numbers are assigned for the hexaquarks and the corresponding mass spectrum is derived.    We have also extrapolated the results both to the physical pion mass and continuum limit.
 
These results are helpful towards the search for such types of exotic states in the future experiments.
 
 \hspace*{\fill} \\

{\it Acknowledgments:}
We thank Zhan-Wei Liu, Zhen-Xing Zhao, Min-Huan Chu, Jun Hua, and Chun-Jiang Shi for useful discussions.  This work is supported in part by Natural Science Foundation of China under grant No. 11735010, 11975127, 11911530088, U2032102, 12005130.  HL, JCH, and WW are also supported  by Natural Science Foundation of Shanghai under grant No. 15DZ2272100.  PS is also supported by Jiangsu Specially Appointed Professor Program. YBY is also supported by the Strategic Priority Research Program of Chinese Academy of Sciences, Grant No. XDB34030303, XDPB15. AS, PS, WW, and YBY are also supported by the NSFC-DFG joint grant under grant No. 12061131006 and SCHA~~458/22. The calculation was supported by the Siyuan Mark 1 cluster at Center for High Performance Computing, Shanghai Jiao Tong University, and National Super computing Center in Zhengzhou.

\appendix

\section{extrapolation of hadrons}

In this appendix, we have collected the masses for the $K, D, D_s, \Lambda, \Xi, \Omega, \Lambda_c, \Xi_c, \eta_c$ and $J/\psi$ on the five ensembles. After the continuum and physical pion  extrapolation, shown in Fig.5, the results for these hadrons are collected in Tab.~\ref{tab:ordinary_meson}. 
\\
\\
\\
\begin{widetext}

\begin{figure}[http]
\centering
\subfigure[~$K$]{\label{fig:subfig:a}
\includegraphics[width=0.35\linewidth]{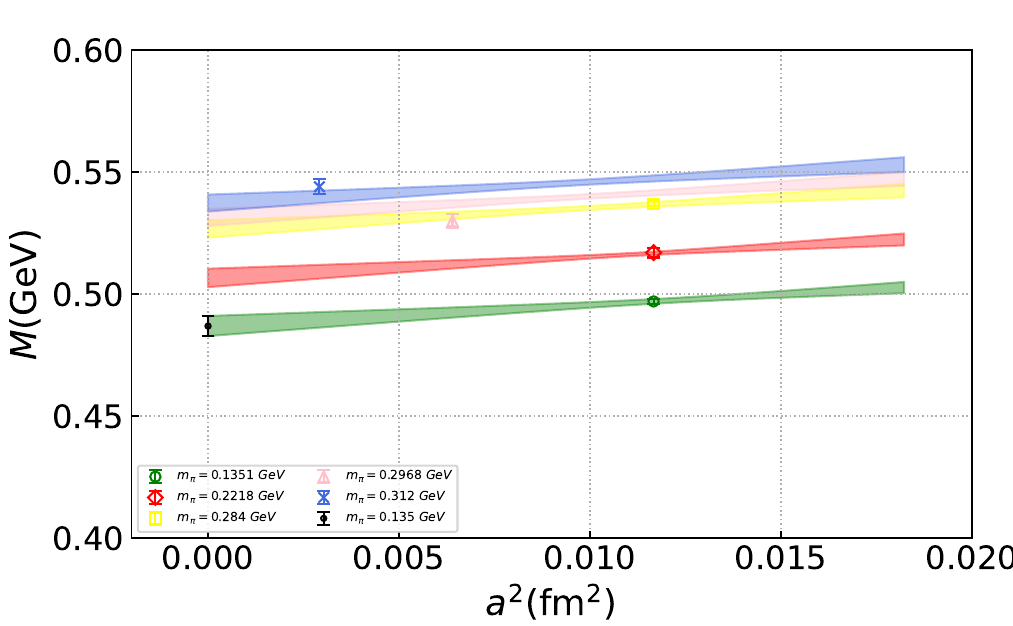}}
\hspace{0.01\linewidth}
\subfigure[~$D$]{\label{fig:subfig:b}
\includegraphics[width=0.35\linewidth]{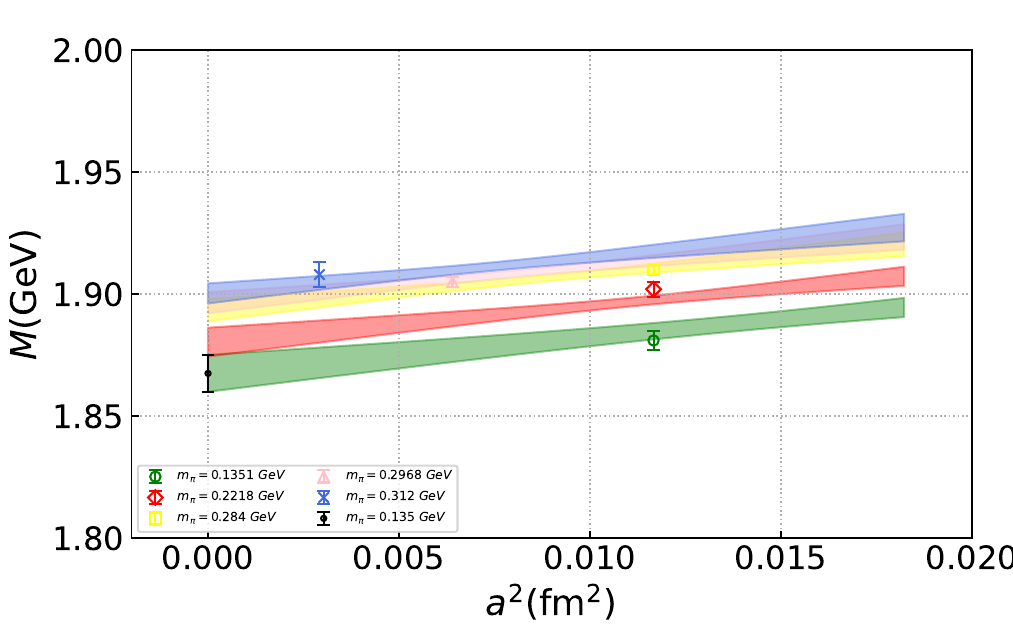}}
\vfill
\subfigure[~$D_s$]{\label{fig:subfig:a}
\includegraphics[width=0.35\linewidth]{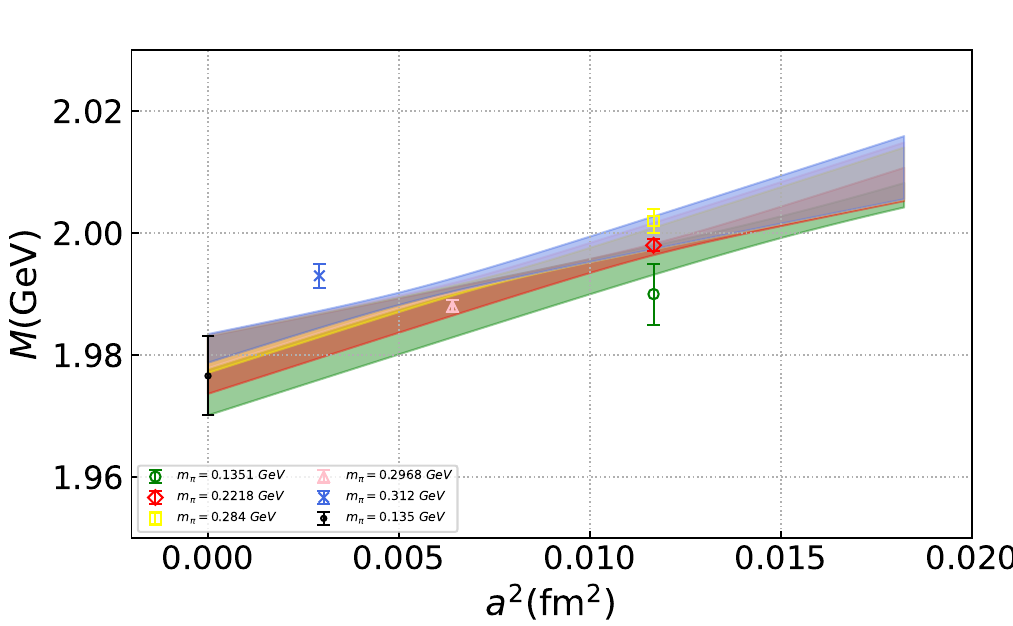}}
\hspace{0.01\linewidth}
\subfigure[~$\Lambda$]{\label{fig:subfig:b}
\includegraphics[width=0.35\linewidth]{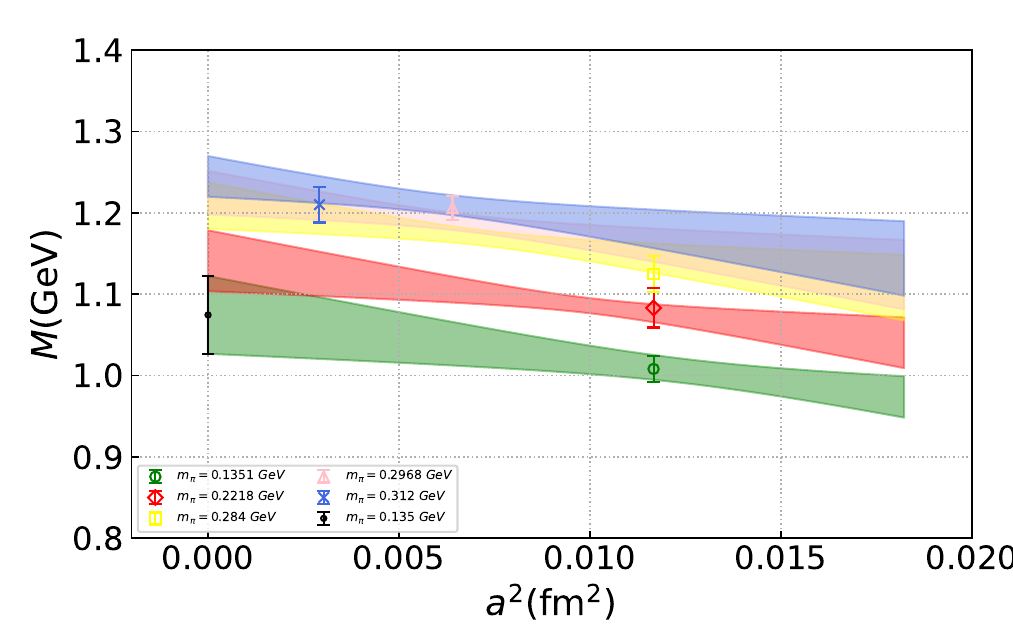}}

\vfill
\subfigure[~$\Xi$]{\label{fig:subfig:a}
\includegraphics[width=0.35\linewidth]{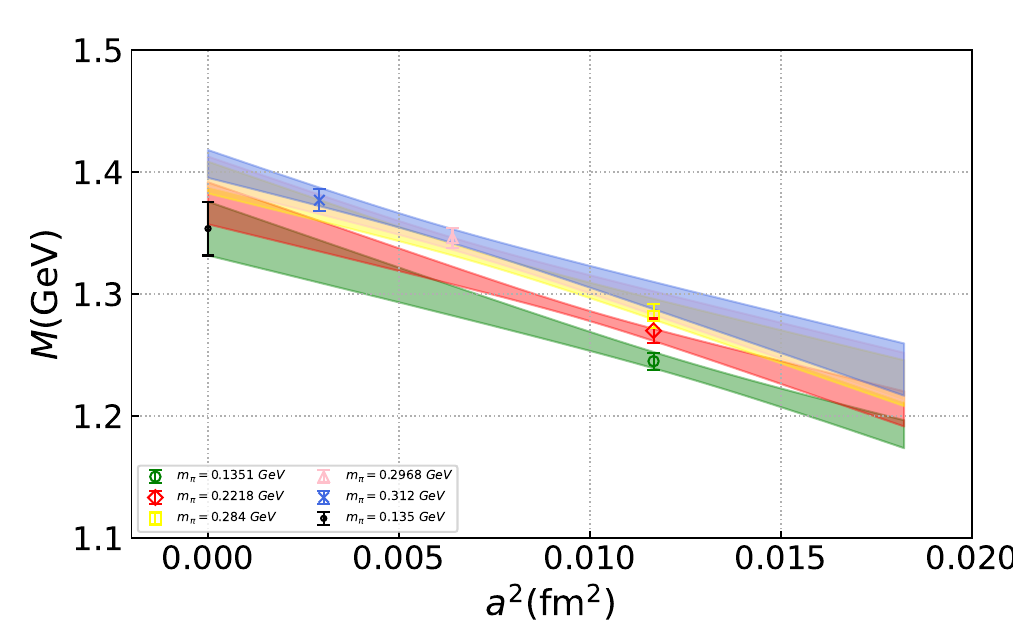}}
\hspace{0.01\linewidth}
\subfigure[~$\Omega$]{\label{fig:subfig:b}
\includegraphics[width=0.35\linewidth]{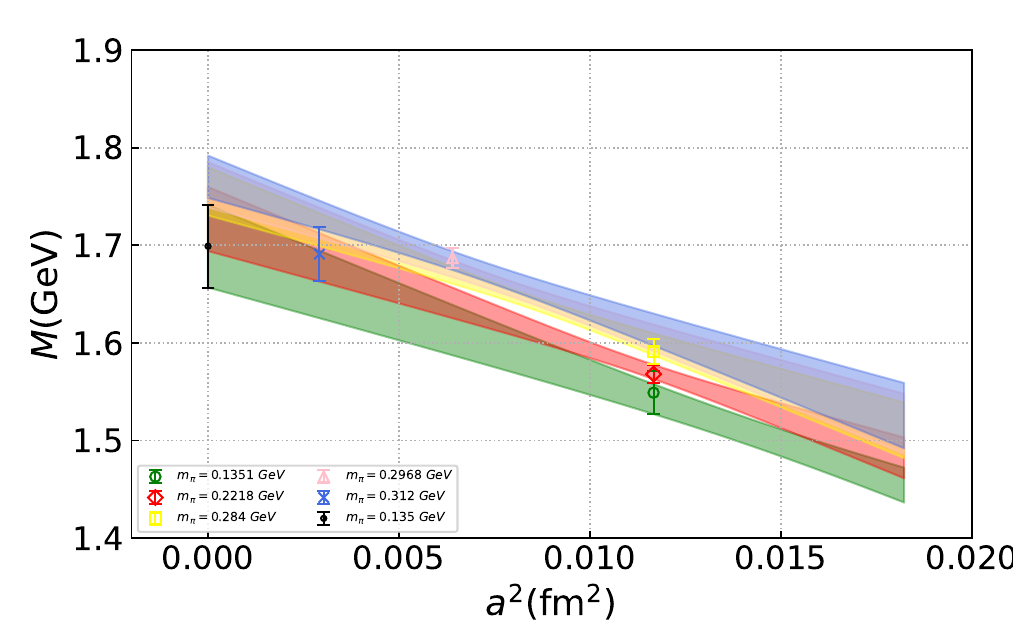}}
\vfill
\subfigure[~$\Lambda_c$]{\label{fig:subfig:a}
\includegraphics[width=0.35\linewidth]{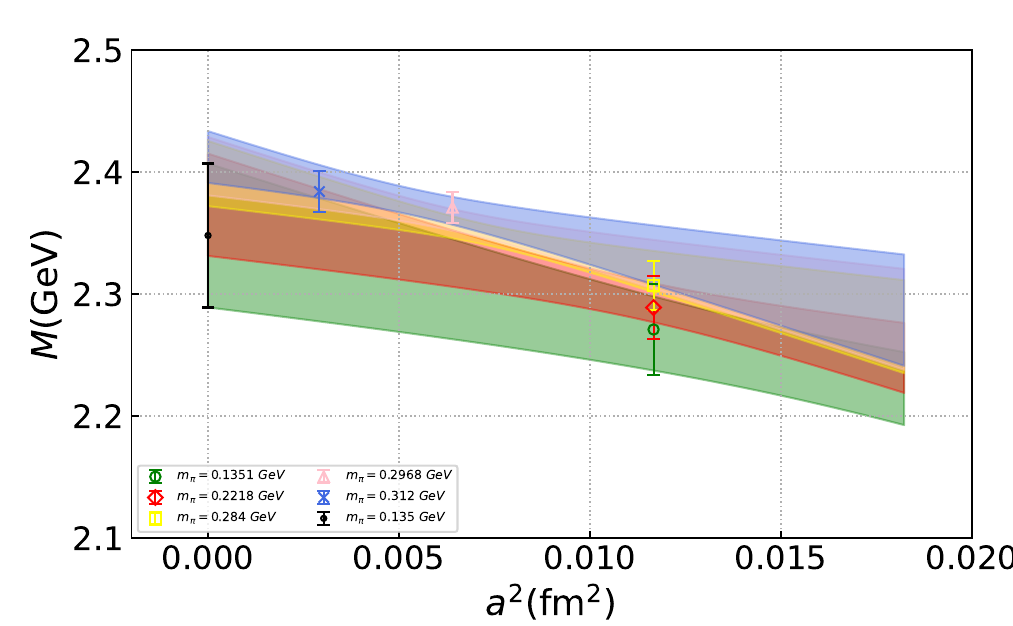}}
\hspace{0.01\linewidth}
\subfigure[~$\Xi_c$]{\label{fig:subfig:b}
\includegraphics[width=0.35\linewidth]{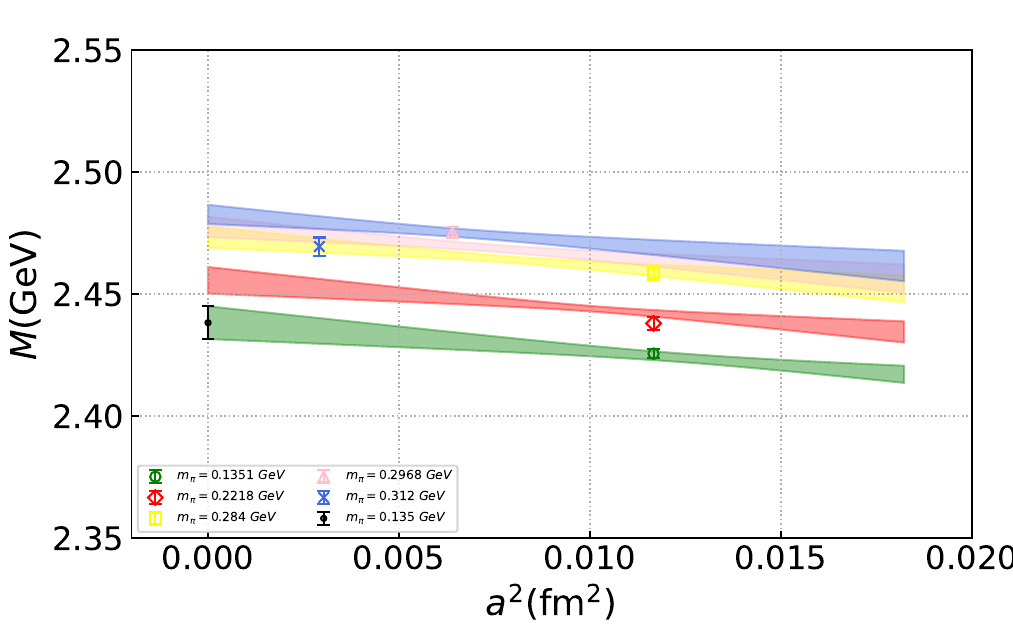}}
\vfill
\subfigure[~$\eta_c$]{\label{fig:subfig:a}
\includegraphics[width=0.35\linewidth]{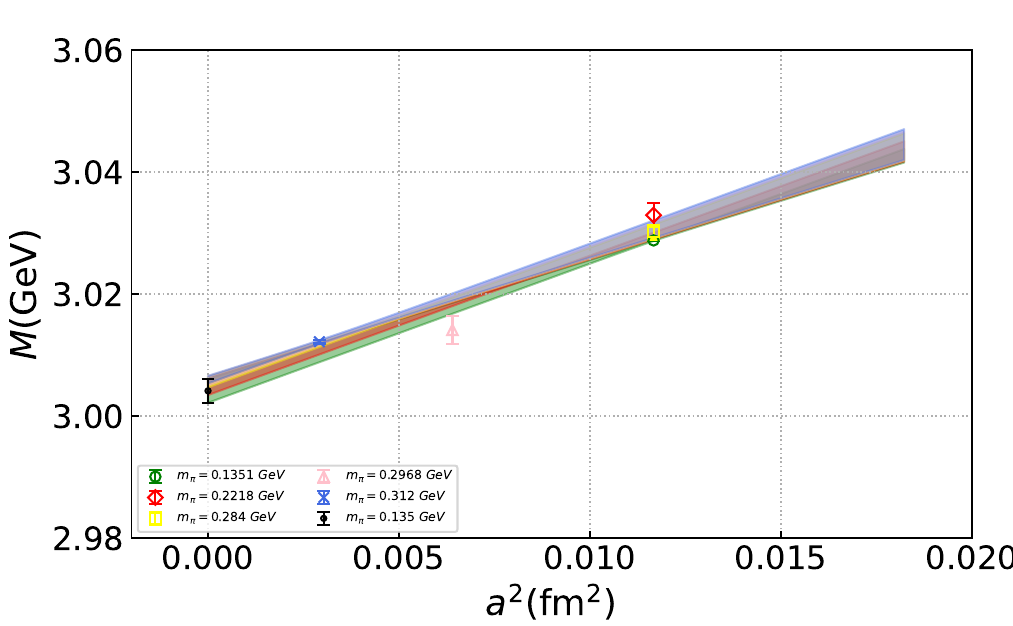}}
\hspace{0.01\linewidth}
\subfigure[~$J/\psi$]{\label{fig:subfig:b}
\includegraphics[width=0.35\linewidth]{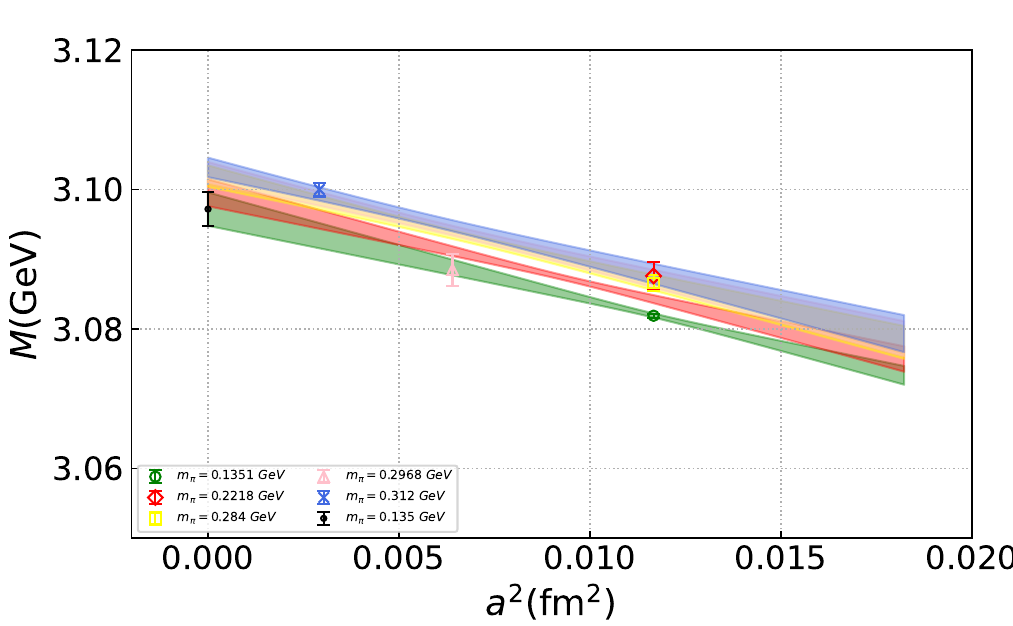}}
\caption{Extrapolation for the mass for the ordinary hadrons: $K$, $D$, $D_s$, $\Lambda$, $\Xi$, $\Omega$, $\Lambda_c$,$\Xi_c$,$\eta_c$ and $J/\psi$.  }
\label{fig:subfig}
\end{figure}

\section{Effective mass on other ensembles}
Effective masses for hexaquarks on the four different quantum numbers on C08P30S, C11P14L, C11P22M, and C11P29S are shown in Fig 6-9.
\begin{figure}[http]
\centering
\subfigure[~$\rm C08P30S$]{\label{fig:subfig:a}
\includegraphics[width=0.35\linewidth]{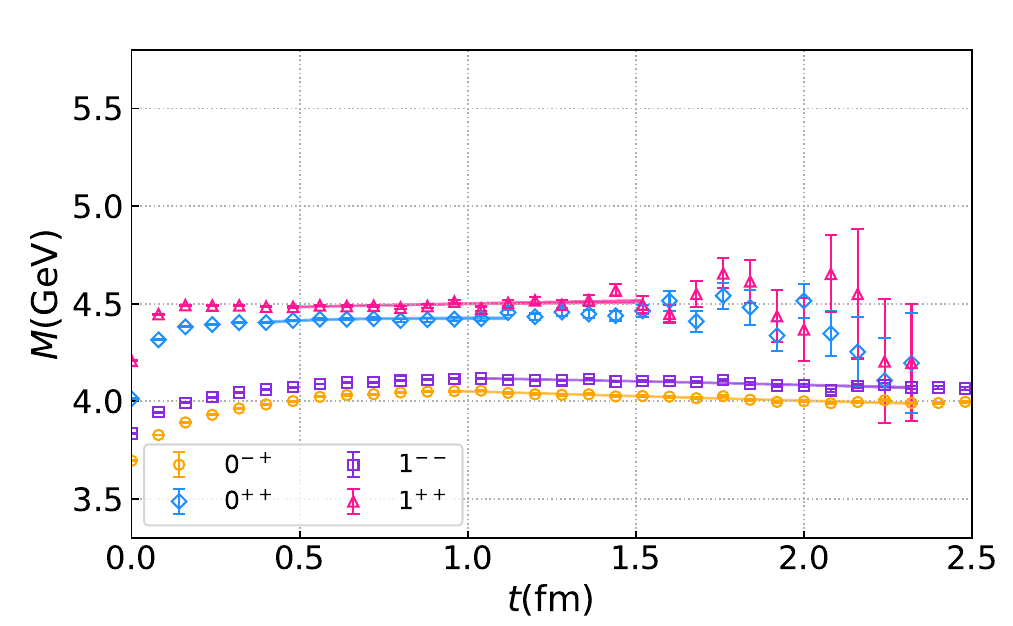}}
\hspace{0.01\linewidth}
\subfigure[~$\rm C11P14L$]{\label{fig:subfig:b}
\includegraphics[width=0.35\linewidth]{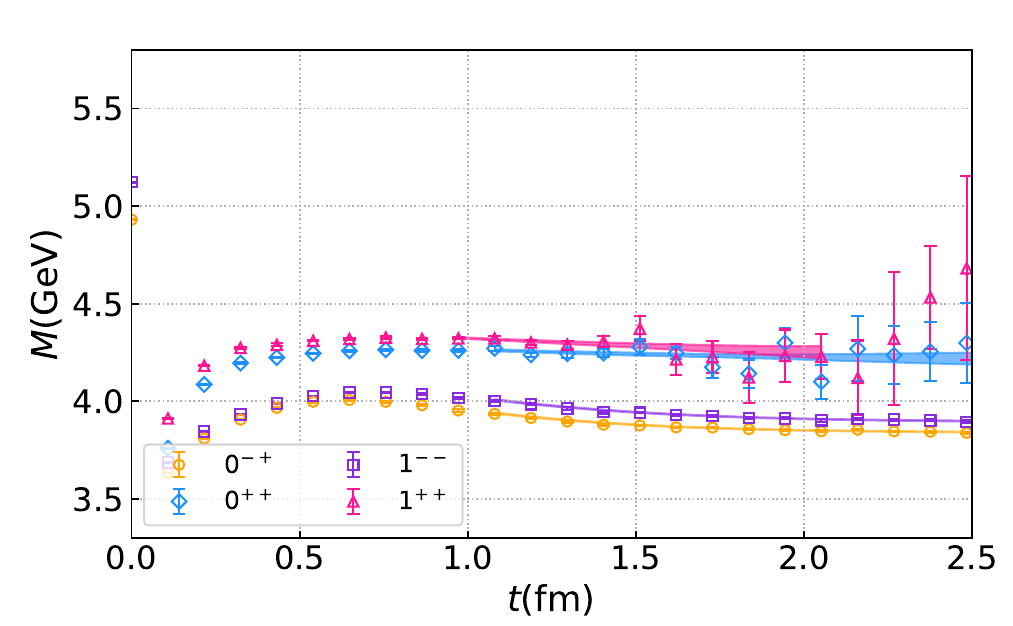}}
\vfill
\subfigure[~$\rm C111P22M$]{\label{fig:subfig:c}
\includegraphics[width=0.35\linewidth]{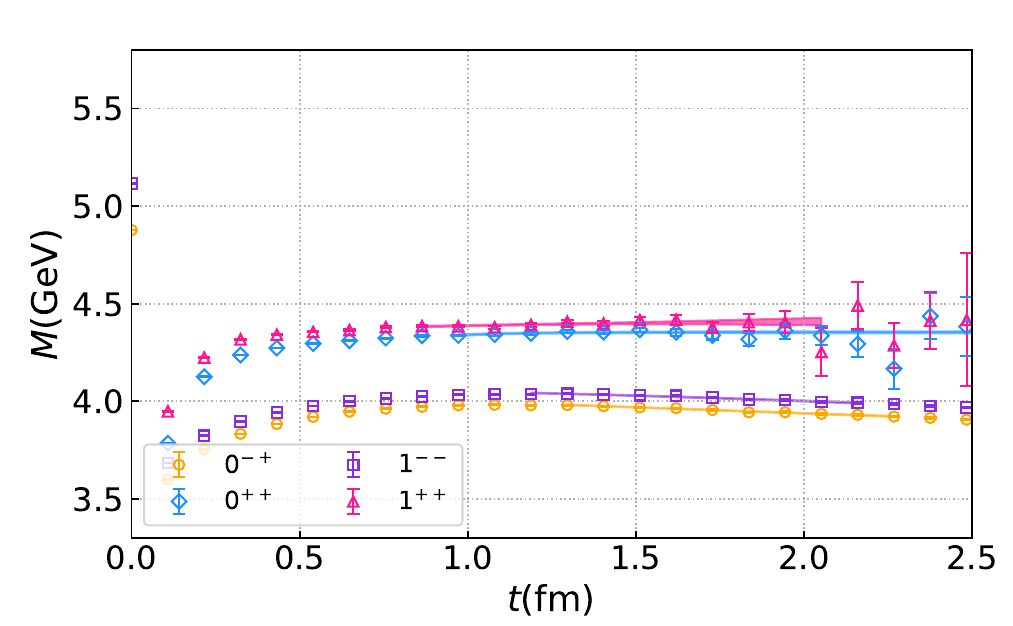}}
\hspace{0.01\linewidth}
\subfigure[~$\rm C11P29S$]{\label{fig:subfig:d}
\includegraphics[width=0.35\linewidth]{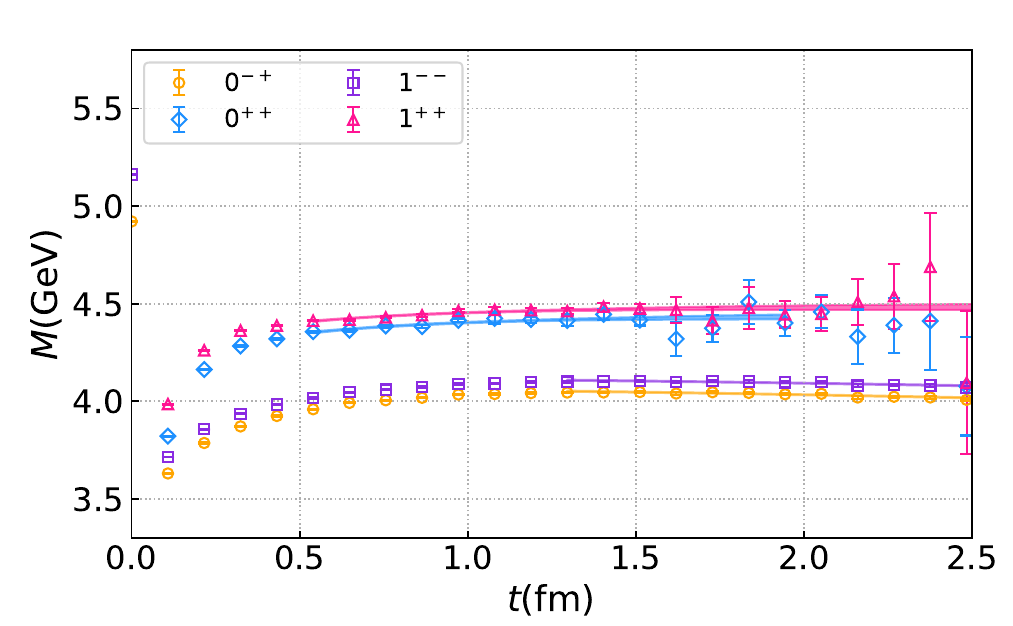}}
\caption{Effective mass of the ground states for hexaquarks on the four 
 different quantum numbers on different ensembles. }
\label{fig:effective_mass}
\end{figure}
\end{widetext}



\begin{thebibliography}{11}


\bibitem{Gell-Mann:2015moa}
M.~Gell-Mann,
doi:10.1142/9789814618113\_0001

\bibitem{Zweig:1964jf}
G.~Zweig,
CERN-TH-412.

\bibitem{Jaffe:1976ig}
R.~L.~Jaffe,
Phys. Rev. D \textbf{15}, 267 (1977)
doi:10.1103/PhysRevD.15.267

\bibitem{Jaffe:1977cv}
R.~L.~Jaffe,
Phys. Rev. D \textbf{17}, 1444 (1978)
doi:10.1103/PhysRevD.17.1444

\bibitem{Lipkin:1987sk}
H.~J.~Lipkin,
Phys. Lett. B \textbf{195}, 484-488 (1987)
doi:10.1016/0370-2693(87)90055-4

\bibitem{Belle:2003nnu}
S.~K.~Choi \textit{et al.} [Belle],
Phys. Rev. Lett. \textbf{91}, 262001 (2003)
doi:10.1103/PhysRevLett.91.262001
[arXiv:hep-ex/0309032 [hep-ex]].

\bibitem{NPLQCD:2012mex}
S.~R.~Beane \textit{et al.} [NPLQCD],
Phys. Rev. D \textbf{87}, no.3, 034506 (2013)
doi:10.1103/PhysRevD.87.034506
[arXiv:1206.5219 [hep-lat]].

\bibitem{CDF:2003cab}
D.~Acosta \textit{et al.} [CDF],
Phys. Rev. Lett. \textbf{93}, 072001 (2004)
doi:10.1103/PhysRevLett.93.072001
[arXiv:hep-ex/0312021 [hep-ex]].

\bibitem{D0:2004zmu}
V.~M.~Abazov \textit{et al.} [D0],
Phys. Rev. Lett. \textbf{93}, 162002 (2004)
doi:10.1103/PhysRevLett.93.162002
[arXiv:hep-ex/0405004 [hep-ex]].

\bibitem{LHCb:2014zfx}
R.~Aaij \textit{et al.} [LHCb],
Phys. Rev. Lett. \textbf{112}, no.22, 222002 (2014)
doi:10.1103/PhysRevLett.112.222002
[arXiv:1404.1903 [hep-ex]].

\bibitem{BESIII:2016adj}
M.~Ablikim \textit{et al.} [BESIII],
Phys. Rev. Lett. \textbf{118}, no.9, 092002 (2017)
doi:10.1103/PhysRevLett.118.092002
[arXiv:1610.07044 [hep-ex]].

\bibitem{BESIII:2016bnd}
M.~Ablikim \textit{et al.} [BESIII],
Phys. Rev. Lett. \textbf{118}, no.9, 092001 (2017)
doi:10.1103/PhysRevLett.118.092001
[arXiv:1611.01317 [hep-ex]].

\bibitem{BESIII:2017tqk}
M.~Ablikim \textit{et al.} [BESIII],
Phys. Rev. D \textbf{96}, no.3, 032004 (2017)
[erratum: Phys. Rev. D \textbf{99}, no.1, 019903 (2019)]
doi:10.1103/PhysRevD.96.032004
[arXiv:1703.08787 [hep-ex]].

\bibitem{BaBar:2005hhc}
B.~Aubert \textit{et al.} [BaBar],
Phys. Rev. Lett. \textbf{95}, 142001 (2005)
doi:10.1103/PhysRevLett.95.142001
[arXiv:hep-ex/0506081 [hep-ex]].

\bibitem{CLEO:2006tct}
Q.~He \textit{et al.} [CLEO],
Phys. Rev. D \textbf{74}, 091104 (2006)
doi:10.1103/PhysRevD.74.091104
[arXiv:hep-ex/0611021 [hep-ex]].

\bibitem{Belle:2007dxy}
C.~Z.~Yuan \textit{et al.} [Belle],
Phys. Rev. Lett. \textbf{99}, 182004 (2007)
doi:10.1103/PhysRevLett.99.182004
[arXiv:0707.2541 [hep-ex]].

\bibitem{Belle:2013yex}
Z.~Q.~Liu \textit{et al.} [Belle],
Phys. Rev. Lett. \textbf{110}, 252002 (2013)
[erratum: Phys. Rev. Lett. \textbf{111}, 019901 (2013)]
doi:10.1103/PhysRevLett.110.252002
[arXiv:1304.0121 [hep-ex]].

\bibitem{BaBar:2006ait}
B.~Aubert \textit{et al.} [BaBar],
Phys. Rev. Lett. \textbf{98}, 212001 (2007)
doi:10.1103/PhysRevLett.98.212001
[arXiv:hep-ex/0610057 [hep-ex]].

\bibitem{BaBar:2012hpr}
J.~P.~Lees \textit{et al.} [BaBar],
Phys. Rev. D \textbf{89}, no.11, 111103 (2014)
doi:10.1103/PhysRevD.89.111103
[arXiv:1211.6271 [hep-ex]].

\bibitem{Belle:2007umv}
X.~L.~Wang \textit{et al.} [Belle],
Phys. Rev. Lett. \textbf{99}, 142002 (2007)
doi:10.1103/PhysRevLett.99.142002
[arXiv:0707.3699 [hep-ex]].

\bibitem{Belle:2014wyt}
X.~L.~Wang \textit{et al.} [Belle],
Phys. Rev. D \textbf{91}, 112007 (2015)
doi:10.1103/PhysRevD.91.112007
[arXiv:1410.7641 [hep-ex]].

\bibitem{BESIII:2020qkh}
M.~Ablikim \textit{et al.} [BESIII],
Phys. Rev. Lett. \textbf{126}, no.10, 102001 (2021)
doi:10.1103/PhysRevLett.126.102001
[arXiv:2011.07855 [hep-ex]].

\bibitem{LHCb:2015yax}
R.~Aaij \textit{et al.} [LHCb],
Phys. Rev. Lett. \textbf{115}, 072001 (2015)
doi:10.1103/PhysRevLett.115.072001
[arXiv:1507.03414 [hep-ex]].

\bibitem{LHCb:2016ztz}
R.~Aaij \textit{et al.} [LHCb],
Phys. Rev. Lett. \textbf{117}, no.8, 082002 (2016)
doi:10.1103/PhysRevLett.117.082002
[arXiv:1604.05708 [hep-ex]].

\bibitem{LHCb:2019kea}
R.~Aaij \textit{et al.} [LHCb],
Phys. Rev. Lett. \textbf{122}, no.22, 222001 (2019)
doi:10.1103/PhysRevLett.122.222001
[arXiv:1904.03947 [hep-ex]].

\bibitem{Dyson:1964xwa}
F.~Dyson and N.~H.~Xuong,
Phys. Rev. Lett. \textbf{13}, no.26, 815-817 (1964)
doi:10.1103/PhysRevLett.13.815

\bibitem{Faldt:2011zv}
G.~Faldt and C.~Wilkin,
Phys. Lett. B \textbf{701}, 619-622 (2011)
doi:10.1016/j.physletb.2011.06.054
[arXiv:1105.4142 [nucl-th]].

\bibitem{WASA-at-COSY:2011bjg}
P.~Adlarson \textit{et al.} [WASA-at-COSY],
Phys. Rev. Lett. \textbf{106}, 242302 (2011)
doi:10.1103/PhysRevLett.106.242302
[arXiv:1104.0123 [nucl-ex]].

\bibitem{WASA-at-COSY:2012seb}
P.~Adlarson \textit{et al.} [WASA-at-COSY],
Phys. Lett. B \textbf{721}, 229-236 (2013)
doi:10.1016/j.physletb.2013.03.019
[arXiv:1212.2881 [nucl-ex]].

\bibitem{Kim:2020rwn}
H.~Kim, K.~S.~Kim and M.~Oka,
Phys. Rev. D \textbf{102}, no.7, 074023 (2020)
doi:10.1103/PhysRevD.102.074023
[arXiv:2009.11983 [hep-ph]].

\bibitem{Dong:2018emq}
Y.~Dong, P.~Shen and Z.~Zhang,
Phys. Rev. D \textbf{97}, no.11, 114002 (2018)
doi:10.1103/PhysRevD.97.114002
[arXiv:1801.04700 [hep-ph]].

\bibitem{Oka:2019mrd}
M.~Oka, S.~Maeda and Y.~R.~Liu,
Int. J. Mod. Phys. Conf. Ser. \textbf{49}, 1960004 (2019)
doi:10.1142/S2010194519600048
[arXiv:1904.00586 [hep-ph]].

\bibitem{Pepin:1998ih}
S.~Pepin and F.~Stancu,
Phys. Rev. D \textbf{57}, 4475-4478 (1998)
doi:10.1103/PhysRevD.57.4475
[arXiv:hep-ph/9710528 [hep-ph]].

\bibitem{Vijande:2016nzk}
J.~Vijande, A.~Valcarce, J.~M.~Richard and P.~Sorba,
Phys. Rev. D \textbf{94}, no.3, 034038 (2016)
doi:10.1103/PhysRevD.94.034038
[arXiv:1608.03982 [hep-ph]].

\bibitem{Meng:2017fwb}
L.~Meng, N.~Li and S.~L.~Zhu,
Phys. Rev. D \textbf{95}, no.11, 114019 (2017)
doi:10.1103/PhysRevD.95.114019
[arXiv:1704.01009 [hep-ph]].

\bibitem{Zhang:1997ny}
Z.~Y.~Zhang, Y.~W.~Yu, P.~N.~Shen, L.~R.~Dai, A.~Faessler and U.~Straub,
Nucl. Phys. A \textbf{625}, 59-70 (1997)
doi:10.1016/S0375-9474(97)00033-X

\bibitem{Gerasyuta:2010hn}
S.~M.~Gerasyuta and E.~E.~Matskevich,
Phys. Rev. D \textbf{82}, 056002 (2010)
doi:10.1103/PhysRevD.82.056002
[arXiv:1003.0257 [hep-ph]].

\bibitem{Park:2015nha}
W.~Park, A.~Park and S.~H.~Lee,
Phys. Rev. D \textbf{92}, no.1, 014037 (2015)
doi:10.1103/PhysRevD.92.014037
[arXiv:1506.01123 [nucl-th]].

\bibitem{Chen:2019vdh}
X.~H.~Chen, Q.~N.~Wang, W.~Chen and H.~X.~Chen,
Chin. Phys. C \textbf{45}, no.4, 041002 (2021)
doi:10.1088/1674-1137/abdfbe
[arXiv:1906.11089 [hep-ph]].

\bibitem{Huang:2020bmb}
H.~Huang, J.~Ping, X.~Zhu and F.~Wang,
[arXiv:2011.00513 [hep-ph]].

\bibitem{Wilson:1974sk}
K.~G.~Wilson,
Phys. Rev. D \textbf{10}, 2445-2459 (1974)
doi:10.1103/PhysRevD.10.2445

\bibitem{Dudek:2007wv}
J.~J.~Dudek, R.~G.~Edwards, N.~Mathur and D.~G.~Richards,
Phys. Rev. D \textbf{77}, 034501 (2008)
doi:10.1103/PhysRevD.77.034501
[arXiv:0707.4162 [hep-lat]].

\bibitem{Dudek:2006ej}
J.~J.~Dudek, R.~G.~Edwards and D.~G.~Richards,
Phys. Rev. D \textbf{73}, 074507 (2006)
doi:10.1103/PhysRevD.73.074507
[arXiv:hep-ph/0601137 [hep-ph]].

\bibitem{Dudek:2009kk}
J.~J.~Dudek, R.~Edwards and C.~E.~Thomas,
Phys. Rev. D \textbf{79}, 094504 (2009)
doi:10.1103/PhysRevD.79.094504
[arXiv:0902.2241 [hep-ph]].

\bibitem{Zhang:2021oja}
Q.~A.~Zhang, J.~Hua, F.~Huang, R.~Li, Y.~Li, C.~D.~Lu, P.~Sun, W.~Sun, W.~Wang and Y.~B.~Yang,
Chin. Phys. C \textbf{46}, no.7, 011002 (2022)
doi:10.1088/1674-1137/ac2b12
[arXiv:2103.07064 [hep-lat]].

\bibitem{Belle:2021crz}
Y.~B.~Li \textit{et al.} [Belle],
Phys. Rev. Lett. \textbf{127}, no.12, 121803 (2021)
doi:10.1103/PhysRevLett.127.121803
[arXiv:2103.06496 [hep-ex]].

\bibitem{Zyla:2020zbs}
P.~A.~Zyla \textit{et al.} [Particle Data Group],
PTEP \textbf{2020}, no.8, 083C01 (2020)
doi:10.1093/ptep/ptaa104

\bibitem{Giusti:2017dmp}
D.~Giusti, V.~Lubicz, C.~Tarantino, G.~Martinelli, F.~Sanfilippo, S.~Simula and N.~Tantalo,
Phys. Rev. D \textbf{95}, no.11, 114504 (2017)
doi:10.1103/PhysRevD.95.114504
[arXiv:1704.06561 [hep-lat]].

\bibitem{Lee:2011rka}
N.~Lee, Z.~G.~Luo, X.~L.~Chen and S.~L.~Zhu,
Phys. Rev. D \textbf{84}, 014031 (2011)
doi:10.1103/PhysRevD.84.014031
[arXiv:1104.4257 [hep-ph]].

\bibitem{Liu:2021gva}
Z.~Liu, H.~T.~An, Z.~W.~Liu and X.~Liu,
Phys. Rev. D \textbf{105}, no.3, 034006 (2022)
doi:10.1103/PhysRevD.105.034006
[arXiv:2112.02510 [hep-ph]].
\bibitem{Dudek:2010wm}
J.~J.~Dudek, R.~G.~Edwards, M.~J.~Peardon, D.~G.~Richards and C.~E.~Thomas,
Phys. Rev. D \textbf{82} (2010), 034508
doi:10.1103/PhysRevD.82.034508
[arXiv:1004.4930 [hep-ph]].
\bibitem{Blossier:2009kd}
B.~Blossier, M.~Della Morte, G.~von Hippel, T.~Mendes and R.~Sommer,
JHEP \textbf{04} (2009), 094
doi:10.1088/1126-6708/2009/04/094
[arXiv:0902.1265 [hep-lat]].
\bibitem{LHCb:2020bwg}
R.~Aaij \textit{et al.} [LHCb],
Sci. Bull. \textbf{65} (2020) no.23, 1983-1993
doi:10.1016/j.scib.2020.08.032
[arXiv:2006.16957 [hep-ex]].
\bibitem{Gasser:1984gg}
J.~Gasser and H.~Leutwyler,
Nucl. Phys. B \textbf{250} (1985), 465-516
doi:10.1016/0550-3213(85)90492-4
\bibitem{CMS:2022yhl}
 [CMS],
CMS-PAS-BPH-21-003.
\bibitem{ATLAS:2022hhx}
 [ATLAS],
ATLAS-CONF-2022-040.

\bibitem{MILC:2010pul}
A.~Bazavov \textit{et al.} [MILC],
Phys. Rev. D \textbf{82} (2010), 074501
doi:10.1103/PhysRevD.82.074501
[arXiv:1004.0342 [hep-lat]].

\bibitem{Koponen:2017ayj}
J.~Koponen, A.~Zimermmane-Santos, C.~Davies, G.~P.~Lepage and A.~Lytle,
EPJ Web Conf. \textbf{175} (2018), 06015
doi:10.1051/epjconf/201817506015
[arXiv:1710.07554 [hep-lat]].

\end{thebibliography}
\end{document}